\documentclass{ws-mpla}

\usepackage[centertags]{amsmath}
\usepackage{amsfonts}
\usepackage{amssymb}
\usepackage{latexsym}
\usepackage[superscript]{cite}

\RequirePackage{hyperref}

\renewcommand{\vec}[1]{\ensuremath{\boldsymbol{\mathrm{#1}}}}
\newcommand{\uI}{\ensuremath{\mathrm{i}}}

\makeatletter
\def\catchline#1#2#3{\expandafter\def\expandafter\@clinebuf\expandafter
    {\@clinebuf\catchlinefont
    \noindent
    \begin{minipage}{\textwidth}
    Published in Modern Physics Letters A, \hfill [\ArXivNo]\\
    Vol.~\textbf{21}, No.~12, pp.~979--984 (2006). \hfill \today\\
    doi:\href{http://dx.doi.org/10.1142/S0217732306019359}%
    {10.1142/S0217732306019359} \hfill \currenttime 
    \end{minipage}
    \hfill 
    \relax\par
    \phantom{A}}\relax\par
    }%
\makeatother \catchline{}{}{}{}{}
\newcommand{\ArXivNo}{\href{http://arxiv.org/abs/hep-th/0605152}{hep-th/0605152}}

\def\draftnote{\today\quad\currenttime\quad [\ArXivNo]\qquad}%
\def\email#1{\upshape\texttt{#1}}

\makeatletter
\let\@afterindentfalse\@afterindenttrue
\@afterindenttrue \makeatother

\allowdisplaybreaks

\begin{document}

\markboth{Nattapong~Yongram, Edouard~B.~Manoukian \&
Suppiya~Siranan} {Polarization Correlations in Muon Pair
Production in the Electroweak Model}

\title{\MakeUppercase{Polarization Correlations in Muon Pair %
Production in the Electroweak Model}\thanks{Work supported by a
``Royal Golden Jubilee Ph.D. Program''.}}

\author{\footnotesize \MakeUppercase{Nattapong~Yongram}, \
\MakeUppercase{Edouard~B.~Manoukian}\thanks{Corresponding author.}
\ and \ \MakeUppercase{Suppiya~Siranan}}

\address{School of Physics, \ Suranaree University of Technology,\\
Nakhon Ratchasima, 30000, Thailand\\
\email{edouard@sut.ac.th}}

\maketitle

\pub{Received 11 October 2005}{Revised 26 October 2005}

\begin{abstract}
  Explicit field theory computations are carried out of the
  joint probabilities associated with spin correlations of
  $\mu^{-}\mu^{+}$ produced in $e^{-}e^{+}$ collision in
  the standard electroweak model to the leading order.
  The derived expressions are found to \emph{depend not only on
  the speed of the $e^{-}e^{+}$ pair but also on the
  underlying couplings}. These expressions are unlike the ones
  obtained from simply combining the spins of the
  relevant particles which are of kinematical nature.
  It is remarkable that these
  explicit results obtained from quantum field theory show a clear
  violation of Bell's inequality.
\end{abstract}

\ccode{\textit{Keywwords}: Polarization correlations, Quantum field
theory, High-Energy computation, The Standard Electroweak Model,
Bell's test}

\ccode{PACS Nos.: 11.15Bt, 12.15Ji, 13.88.+e, 03.65.Ud }

\section*{{}}

Several experiments have been performed over the years on
particles' polarizations
correlations~\cite{Irby_2003,Osuch_1996,Kaday_1975,Fry_1995,Aspect_1982}
in the light of Bell's inequality and many Bell-like experiments
have been proposed recently in high energy
physics.~\cite{Go_2004,Bertlmann_2004,Abel_1992,Privitera_1992,Lednicky_2001,Genovese_2001}
We have been particularly interested in actual quantum field
theory computations of polarizations correlations probabilities of
particles produced in basic processes because of novelties
encountered in dynamical calculations as opposed to kinematical
considerations to be discussed. Here it is worth recalling that
quantum field theory originates from the combination of quantum
physics \emph{and} relativity and involve non-trivial dynamics.
Many such computations have been done in
QED~\cite{Yongram_2003,Manoukian_2004} as well as in $e^{-}e^{+}$
pair production from some charged and neutral
strings~\cite{Manoukian_2005}.  All of these polarizations
correlations probabilities based on dynamical analyses following
from field theory share the interesting property that they depend
on the energy (speeds) of the colliding particles due to the mere
fact that typically the latter carry speeds in order to collide.
Such analyses are unlike considerations based on formal arguments
of simply combining spins~\cite{Clauser_1978}, as is usually done,
and are of kinematical nature, void of dynamical considerations.
Here it is worth recalling that the total spin of a two-particle
system each with spin, such as of two spin~$1/2$'s, is obtained
not only from combining the spins of the latter but also from any
orbital angular momentum residing in their center of mass system.
For low speeds, one expects that the argument based simply on
combining the spins of the colliding particles should provide an
accurate description of the polarization correlations sought and
all of our QED computations~\cite{Yongram_2003,Manoukian_2004}
show the correctness of such an argument in the limit of low
speeds. Needless to say, we are interested in the
\emph{relativistic} regime as well, and the formal arguments just
mentioned fail to provide the correct expressions for the
correlations. As a byproduct of the work, our computations of the
joint polarizations correlations carried out in a full quantum
field theory setting show a clear violation of Bell's inequality.

In the present communication we encounter additional
\emph{completely novel properties} not encountered in our earlier
QED~\cite{Yongram_2003,Manoukian_2004} calculations. We consider
the process $e^{-}e^{+}\to\mu^{-}\mu^{+}$ as described in the
standard electroweak (EW) model. It is well known that this
process~\cite{Althoff_1984} as computed in the EW model is in much
better agreement with experiments than that of a QED computation.
The reasons for considering such a process in the EW model are
many, one of which is the high precision of the differential cross
section obtained as just discussed. Reasons which are, however,
more directly relevant to our anylyses are the following.
\emph{Due to the theshold energy needed to create the
$\mu^{-}\mu^{+}$ pair, the limit of the speed $\beta$ of the
colliding particles cannot be taken to go to zero}. This is unlike
processes treated by the authors in QED such as in
$e^{-}e^{-}\to{}e^{-}e^{-}$, $e^{+}e^{-}\to2\gamma$, Therefore all
arguments based simply on combining the spins of $e^{-}$, $e^{+}$,
without dynamical considerations, \emph{fail}. [As a matter of
fact the latter argument would lead for the joint probability
\emph{in} (\ref{Eq7}) we are seeking, the incorrect result
$(1/2)\sin^{2}\left(\left(\chi_{1}-\chi_{2}\right)/2\right)$---an
expression which has been used for years.] Another novelty we
encounter in the present investigation is that the polarization
correlations \emph{not only depend on speed but have also an
explicit dependence on the underlying couplings}. Again this
latter explicit dependence is unlike the situation arising in
QED~\cite{Yongram_2003,Manoukian_2004}.

The relevant quantity of interest here in testing Bell's
inequality~\cite{Clauser_1974,Clauser_1978} is, in a standard
notation,
\begin{align}
  S &= \frac{p_{12}(a_{1},a_{2})}{p_{12}(\infty,\infty)}
  -\frac{p_{12}(a_{1},a'_{2})}{p_{12}(\infty,\infty)}
  +\frac{p_{12}(a'_{1},a_{2})}{p_{12}(\infty,\infty)}
  +\frac{p_{12}(a'_{1},a'_{2})}{p_{12}(\infty,\infty)}
  \nonumber \\[0.5\baselineskip]
  &\quad
  -\frac{p_{12}(a'_{1},\infty)}{p_{12}(\infty,\infty)}
  -\frac{p_{12}(\infty,a_{2})}{p_{12}(\infty,\infty)}
  \label{Eq1}
\end{align}
as is \emph{computed from} the electroweak model.   Here $a_{1}$,
$a_{2}$ \ $(a'_{1},a'_{2})$ specify directions along which the
polarizations of two particles are measured, with
$p_{12}(a_{1},a_{2})/p_{12}(\infty,\infty)$ denoting the joint
probability, and $p_{12}(a_{1},\infty)/p_{12}(\infty,\infty)$,\
$p_{12}(\infty,a_{2})/p_{12}(\infty,\infty)$ denoting the
probabilities when the polarization of only one of the particles
is measured.   [$p_{12}(\infty,\infty)$ is a normalization
factor.] The corresponding probabilities as computed from the
electroweak model will be denoted by $P(\chi_{1},\chi_{2})$,
$P(\chi_{1},-)$, $P(-,\chi_{2})$ with $\chi_{1}$, $\chi_{2}$
denoting angles specifying directions along which spin
measurements are carried out with respect to certain axes spelled
out in the bulk of the paper. To show that the electroweak model
is in violation with Bell's inequality of LHV, it is sufficient to
find one set of angles $\chi_{1}$, $\chi_{2}$, $\chi'_{1}$,
$\chi'_{2}$, such that $S$, as computed in the electroweak model,
leads to a value of $S$ outside the interval $[-1,0]$. In this
work, it is implicitly assumed that the polarization parameters in
the particle states are directly observable and may be used for
Bell-type measurements as discussed.

\begin{figure}[hb!]
  \centering
  \includegraphics[width=0.65\textwidth]{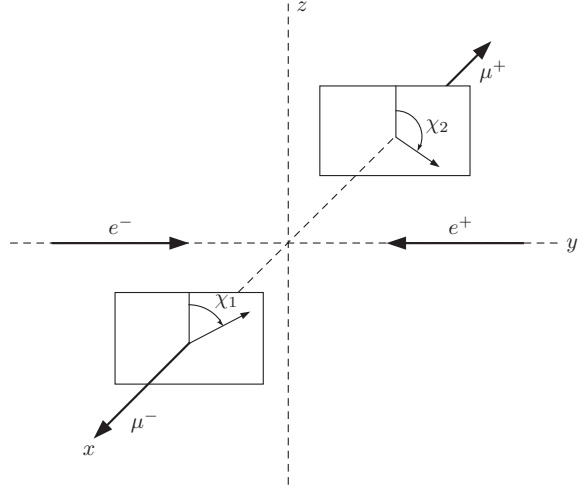}
  \caption{The figure depicts the process
  $e^{-}e^{+}\to\mu^{-}\mu^{+}$, with $e^{-}$, $e^{+}$
  moving along the $y$--axis, and the emerging muons moving along
  the $x$--axis.   $\chi_{1}$ and $\chi_{2}$ denote the angles
  with the $z$--axis specifying the directions of measurements of
  the spins of $\mu^{-}$ and $\mu^{+}$, respectively.}
  \label{Fig1}
\end{figure}

We consider the process $e^{-}e^{+}\to\mu^{-}\mu^{+}$ in the
center of mass frame (see figure~\ref{Fig1}) with the momentum of,
say, $e^{-}$ chosen to be
$\vec{p}=\gamma\beta{}m_{e}\big(0,1,0\big)={-}\vec{k}$,\ $m_{e}$
denoting its mass and $\gamma=1/\sqrt{1-\beta^{2}}$.   The
momentum of the emerging $\mu^{-}$ will be taken to be $\vec{p}' =
\gamma'\beta'm_{\mu}\big(1,0,0\big) = {-}\vec{k}'$,
$\gamma'=1/\sqrt{1-\beta'^{2}}$,\ and $m_{\mu}$ is the mass of
$\mu^{-}(\mu^{+})$, the spinors of $e^{-}$, $e^{+}$ are chosen as
\begin{equation}\label{Eq2}
  u(p) = \sqrt{\frac{\gamma+1}{2}}
  \begin{pmatrix}
  \uparrow \\ \uI\frac{\gamma\beta}{\gamma+1}\downarrow
  \end{pmatrix}
  \quad\text{and}\quad
  v(k) = \sqrt{\frac{\gamma+1}{2}}
  \begin{pmatrix}
  \uI\frac{\gamma\beta}{\gamma+1}\uparrow \\ \downarrow
  \end{pmatrix}.
\end{equation}

Obviously, there is a non-zero probability of occurrence of the
above process.  Given that such a process has occurred, we compute
the conditional joint probability of spins measurements of
$\mu^{-}$, $\mu^{+}$ along directions specified by the angles
$\chi_{1}$, $\chi_{2}$ as shown in figure~\ref{Fig1}. Here we have
considered the so-called singlet state. The triplet state leads to
an expression similar to the one in (\ref{Eq7}) for the probability
in question with different coefficients $A(\mathcal{E})$, $\ldots$,
$E(\mathcal{E})$, $N(\mathcal{E})$ and leads again to a violation of
Bell's inequality. The corresponding details may be obtained from
the authors by the interested reader.

A fairly tedious computation for the invariant amplitude of the
process~\cite{Commins_1983,Greiner_1996,Renton_1990} in
figure~\ref{Fig1} leads to
\begin{align}
  \mathcal{M} &\propto \left[A(\mathcal{E})\sin\left(\frac{\chi_{1}-\chi_{2}}{2}\right)
  +B(\mathcal{E})\sin\left(\frac{\chi_{1}+\chi_{2}}{2}\right)
  +C(\mathcal{E})\cos\left(\frac{\chi_{1}-\chi_{2}}{2}\right)\right]
  \nonumber \\[0.5\baselineskip]
  &\qquad -\uI\left[D(\mathcal{E})\sin\left(\frac{\chi_{1}+\chi_{2}}{2}\right)
  +E(\mathcal{E})\cos\left(\frac{\chi_{1}-\chi_{2}}{2}\right)
  \right]
  \label{Eq3}
\end{align}
where
\begin{subequations}
\begin{align}
  A(\mathcal{E}) &= \left(\frac{M^{2}_{Z}}{4\mathcal{E}^{2}}+ab^{2}-1\right)
  \label{Eq4a} \\[0.5\baselineskip]
  B(\mathcal{E}) &= {-}\left(\frac{m_{e}}{m_{\mu}}\right)
  \left(\frac{M^{2}_{Z}}{4\mathcal{E}^{2}}+ab^{2}-1\right)
  \label{Eq4b} \\[0.5\baselineskip]
  C(\mathcal{E}) &=
  \frac{abm_{e}}{\mathcal{E}m_{\mu}}\sqrt{\mathcal{E}^{2}-m^{2}_{\mu}}
  \label{Eq4c} \\[0.5\baselineskip]
  D(\mathcal{E}) &= \frac{a}{m_{\mu}\mathcal{E}}\sqrt{\mathcal{E}^{2}-m^{2}_{\mu}}
  \:\sqrt{\mathcal{E}^{2}-m^{2}_{e}}
  \label{Eq4d} \\[0.5\baselineskip]
  E(\mathcal{E}) &= {-}\frac{ab}{m_{\mu}}\sqrt{\mathcal{E}^{2}-m^{2}_{e}}
  \label{Eq4e}
\end{align}
and
\begin{equation}\label{Eq4f}
  a \equiv \dfrac{g^{2}}{16e^{2}\cos^{2}\theta_{\mathrm{W}}}
  \cong{} 0.353,\qquad
  b \equiv 1-4\sin^{2}\theta_{\mathrm{W}}
  \cong{} 0.08
\end{equation}
\end{subequations}
$g$ denotes the weak coupling constant, $\theta_{\mathrm{W}}$ is
the Weinberg angle, and $e$ denotes the electric charge.  The
contribution of the Higgs particles turns out to be too small and
is negligible.~\cite{Greiner_1996}

Using the notation $F(\chi_{1},\chi_{2})$ for the absolute value
squared of the right-hand side of (\ref{Eq3}), the conditional
joint probability distribution of spin measurements along the
directions specified by the angles $\chi_{1}$, $\chi_{2}$ is given
by
\begin{equation}\label{Eq5}
  P(\chi_{1},\chi_{2}) = \frac{F(\chi_{1},\chi_{2})}{N(\mathcal{E})}
\end{equation}
where the normalization factor $N(\mathcal{E})$ is
\begin{align}
  N(\mathcal{E}) &\equiv{} F(\chi_{1},\chi_{2})+F(\chi_{1}+\pi,\chi_{2})+
  F(\chi_{1},\chi_{2}+\pi)+F(\chi_{1}+\pi,\chi_{2}+\pi)
  \nonumber \\[0.5\baselineskip]
  &= 2\Big\{\big[A(\mathcal{E})\big]^{2}
  +\big[B(\mathcal{E})\big]^{2}+\big[C(\mathcal{E})\big]^{2}
  +\big[D(\mathcal{E})\big]^{2}+\big[E(\mathcal{E})\big]^{2}\Big\}
  \label{Eq6}
\end{align}
giving
\begin{align}
  P(\chi_{1},\chi_{2}) &= \frac{1}{N(\mathcal{E})}
  \left[A(\mathcal{E})\sin\left(
  \frac{\chi_{1}-\chi_{2}}{2}\right)+B(\mathcal{E})
  \sin\left(\frac{\chi_{1}+\chi_{2}}{2}\right)
  +C(\mathcal{E})\cos\left(\frac{\chi_{1}-\chi_{2}}{2}
  \right)\right]^{2}
  \nonumber \\[0.5\baselineskip]
  &\quad +\frac{1}{N(\mathcal{E})}
  \left[D(\mathcal{E})
  \sin\left(\frac{\chi_{1}+\chi_{2}}{2}\right)
  +E(\mathcal{E})\cos\left(\frac{\chi_{1}-\chi_{2}}{2}\right)\right]^{2}.
  \label{Eq7}
\end{align}

The probabilities associated with the measurement of only one of
the polarizations are given respectively, by
\begin{equation}\label{Eq8}
  P(\chi_{1},-) =
  \frac{1}{2}-\frac{2B(\mathcal{E})}{N(\mathcal{E})}
  \big[A(\mathcal{E})\cos\chi_{1}+C(\mathcal{E})\sin\chi_{1}\big]
\end{equation}
and similarly for $\chi_{2}$
\begin{equation}\label{Eq9}
  P(-,\chi_{2}) =
  \frac{1}{2}+\frac{2B(\mathcal{E})}{N(\mathcal{E})}
  \big[A(\mathcal{E})\cos\chi_{2}+C(\mathcal{E})\sin\chi_{2}\big].
\end{equation}

It is important to note that
$P(\chi_{1},\chi_{2})\neq{}P(\chi_{1},-)P(-,\chi_{2})$, in
general, showing the obvious correlations occurring between the
two spins.

The indicator $S$ in (\ref{Eq1}) computed according to the
probabilities $P(\chi_{1},\chi_{2})$, $P(\chi_{1},-)$,
$P(-,\chi_{2})$ in (\ref{Eq7}), (\ref{Eq8}), (\ref{Eq9}) may be
readily evaluated. To show violation of Bell's inequality, it is
sufficient to find four angles $\chi_1$, $\chi_2$, $\chi'_1$,
$\chi'_2$ at accessible energies, for which $S$ falls outside the
interval $[{-}1,0]$. For $\mathcal{E}=105.656$~MeV, i.e., near
threshold, an optimal value of $S$ is obtained equal to $-1.28203$,
for $\chi_{1}=0^{\circ}$, $\chi_{2}=45^{\circ}$,
$\chi'_{1}=90^{\circ}$, $\chi'_{2}=135^{\circ}$, clearly violating
Bell's inequality. For the energies originally carried out in the
experiment on the differential cross section at
$\mathcal{E}\sim34$~GeV, an optimal value of $S$ is obtained equal
to $-1.22094$ for $\chi_{1}=0^{\circ}$, $\chi_{2}=45^{\circ}$,
$\chi'_{1}=51.13^{\circ}$, $\chi'_{2}=170.85^{\circ}$.

As mentioned in the introductory part of the paper, one of the
reasons for this investigation arose from the fact that the limit
of the speed $\beta$ of $e^{-}e^{+}$ cannot be taken to go to zero
due to the threshold energy needed to create the $\mu^{-}\mu^{+}$
pair and methods used for years by simply combining the spins of
the particles in question completely fail. The present
computations are expected to be relevant near the threshold energy
for measuring the spins of the $\mu^{-}\mu^{+}$ pair. Near the
threshold, the indicator $S_{\mathrm{QED}}$ computed within QED
coincides with that of $S$ given above in the electroweak model,
and varies slightly at higher energies, thus confirming that the
weak effects are negligible. Due to the persistence of the
dependence of the indicator $S$ on speed, as seen above, in a
non-trivial way, it would be interesting if any experiments may be
carried out to assess the accuracy of the indicator $S$ as
computed within (relativistic) quantum field theory.  As there is
ample support of the dependence of polarizations correlations, as
we have shown by explicit computations in quantum field theory in
the electroweak interaction as well as QED
ones,~\cite{Yongram_2003,Manoukian_2004} on speed, we hope that
some new experiments will be carried out in the light of Bell-like
tests which monitor speed as further practical tests of quantum
physics in the relativistic regime.

\section*{Acknowledgments}

The authors would like to acknowledge with thanks for being
granted a ``Royal Golden Jubilee Ph.D. Program'' by the Thailand
Research Fund (Grant~No. PHD/0022/2545) for especially carrying
out this project.

\end{document}